\documentstyle[epsfig]{elsart}

\begin{document}
\begin{frontmatter}
\title{The associated photoproduction of positive kaons and
$\pi^0\, \Lambda$ or $\pi\, \Sigma$ pairs in the region of the
$\Sigma$(1385) and $\Lambda$(1405) resonances}

\author{Matthias F. M. Lutz}
\address{GSI, Planckstrasse 1,  D-64291 Darmstadt, Germany \\
Institut f\"ur Kernphysik, TU Darmstadt,
   D-64289 Darmstadt, Germany}
\author{Madeleine Soyeur}
\address{D\'{e}partement d'Astrophysique, de Physique des Particules,\\
de Physique Nucl\'{e}aire et de l'Instrumentation Associ\'{e}e,\\
Service de Physique Nucl\'{e}aire,\\
Commissariat \`{a} l'Energie Atomique/Saclay,
\\F-91191 Gif-sur-Yvette Cedex, France}

\begin{abstract}

The $\gamma p \rightarrow K^+ \pi^0 \Lambda$ and $\gamma p \rightarrow K^+ \pi \Sigma $
reactions are studied in the kinematic region where the $\pi^0 \Lambda$(1116) and $\pi\Sigma$(1192) pairs
originate dominantly from the decay of the $\Sigma$(1385) and $\Lambda$(1405) resonances.
We consider laboratory photon energies around 2 GeV, significantly above the threshold
for producing the $K^+\,\Sigma$(1385)
and $K^+\,\Lambda$(1405)
final states.
We compute for both reactions
the process in which the ingoing photon dissociates
into a real K$^+$ and a virtual K$^-$, the off-shell K$^-$ scattering subsequently
off the proton target to produce the $\pi^0\, \Lambda$ or $\pi\, \Sigma$ pair.
The $K^- p \rightarrow \pi^0 \Lambda$ and $K^- p \rightarrow \pi \Sigma$
amplitudes are calculated in the framework of a chiral coupled-channel
effective field theory of meson-baryon scattering. The structure of the amplitudes
reflects the dominance of the $\Lambda(1405)$ in the $\pi \Sigma$ channel
and of the $\Sigma(1385)$ in the $\pi \Lambda$ channel. The full pion-hyperon
final state interaction is included in these amplitudes.
We extract from the calculated cross section the gauge-invariant double kaon pole
term. We found this term to be large and leading to sizeable cross sections
for both the
$\gamma p \rightarrow K^+ \pi^0 \Lambda$ and $\gamma p \rightarrow K^+ \pi \Sigma $
reactions,
in qualitative agreement with the scarce data presently available.
Accurate measurements of these cross sections
should make it possible to extract the contribution of the double kaon pole
and hence to assess the possibility of studying kaon-nucleon dynamics
just below threshold through these reactions.

\vskip 0.3truecm


\end{abstract}

\end{frontmatter}

\newpage
\section{Introduction}

The simplest process leading to strange particle creation in photon-nucleon
interactions is the associated production of a kaon and a hyperon. The hyperon decays
subsequently into specific channels. We consider
the interaction of 2~ GeV photons (in the laboratory reference frame) with proton targets.
We select final states consisting
of a K$^+$ and a $\pi^0\, \Lambda$ or a $\pi\, \Sigma$ pair respectively and restrict
the invariant mass of these pairs to the mass range of the $\Sigma^0(1385)$
and the $\Lambda(1405)$. The $\Sigma^0(1385)$ decays primarily into the
$\pi^0\, \Lambda$ channel ($88\pm2\,\%$) and less importantly ($12\pm2\,\%$)
into the $\pi\, \Sigma$ channel \cite{Hagiwara}. The $\Lambda(1405)$ decays
entirely into the $\pi\, \Sigma$ channel \cite{Hagiwara}. The $\Sigma^0(1385)$
and the $\Lambda(1405)$ are therefore expected to dominate the production of
$\pi^0\, \Lambda$ and $\pi\, \Sigma$ pairs in the correspon\-ding mass range.

These resonances overlap in mass and have to be separated
experimentally by distinctive decays. A particularly
interesting idea is to study the $\Lambda(1405)$ in the $\pi^0\, \Sigma^0$ channel
\cite{Schmieden}. The transition between the $\Sigma^0(1385)$ and the
$\pi^0\, \Sigma^0$ channel is forbidden because the isospin
Clebsch-Gordan coefficient vanishes. The $\pi^0\, \Sigma^0$ decay is
therefore a unique signature of the $\Lambda(1405)$ channel. Such a
measurement has not yet been performed but is intended at
ELSA (Bonn) where the $\pi^0\, \Sigma^0$ pair could be detected through
a multi-photon final state ($\pi^0\, \Sigma^0 \rightarrow \pi^0\,\Lambda(1116)\,\gamma \rightarrow
 \pi^0\,n \,\pi^0\,\gamma$) with the Crystal Barrel \cite{Schmieden}.
The photoproduction of  $\Lambda(1405)$ resonances is also presently
studied in the charged decay channels, $\pi^-\, \Sigma^+$ and $\pi^+\, \Sigma^-$,
at SPring-8/LEPS with incident photon
energies in the range 1.5 $<E_\gamma^{Lab}<2.4$ GeV \cite{Ahn} and at ELSA with the SAPHIR detector
at 2.6 GeV \cite{SAPHIR}.
These channels are expected to be do\-minated
by the $\Lambda(1405)$ with some contribution from the $\Sigma^0(1385)$.
There are no published cross sections yet.
The only data presently available in the kinematics of interest \cite{Azemoon}
were obtained at DESY thirty years ago with space-like photons, in electroproduction
experiments where the scattered electron and the produced K$^+$ are detected
in coincidence. With this method the
$\Sigma^0(1385)$ and $\Lambda(1405)$ channel could not
be separated at all. The differential cross sections for the $e\, p\rightarrow e\, K^+\,Y$
reaction obtained in these measurements characterize globally strangeness production processes
for missing masses ranging from 1.35 GeV until 1.45 GeV. An interesting
trend of these data is that the t-dependence  of the cross section, for given
photon energy and virtuality, seems to show a sharp drop as would be expected
if the dynamics were dominated by t-channel exchanges. The Mandelstam variable t
is defined as the square of the
4-momentum transfer from the proton target to the $\Sigma^0(1385)$ or $\Lambda(1405)$.

Both the $\Sigma(1385)$ and $\Lambda(1405)$ resonances are located close and below
the $\bar K N$ threshold. These resonances seem to be of rather different nature.
The $\Sigma(1385)$ belongs to the large N$_c$ ground state baryons
and appears well-described by quark models \cite{Isgur,Glozman}. The $\Lambda(1405)$
is a complex baryonic state. Its mass, in particular the large
splitting between the $\Lambda_{1/2^-}(1405)$ and the $\Lambda_{3/2^-}(1520)$,
cannot be understood in the constituent quark model with residual quark-quark interactions
fitting the other low-lying baryonic states \cite{Isgur,Glozman}. There are many
indications that the quark model description of the $\Lambda(1405)$,
if valid at all, requires
a sizeable $q^4\bar q$
component \cite{Jaffe,deSwart,Kaxiras}. This observation is closely related
to the early idea that the $\Lambda(1405)$ can be viewed as a bound kaon-nucleon system
\cite{Dalitz1,Ball,Arnold,Wyld1,Dalitz2,Logan,Rajasekaran,Siegel} and to the later picture
of the $\Lambda(1405)$ as a kaon-soliton bound state \cite{Callan,Blom}.
The $\bar K N$ nature of the $\Lambda(1405)$ was also inferred from the
SU(3) cloudy bag model description \cite{Veit1,Veit2}. Extensive studies
of the $\Lambda(1405)$ based on chiral Lagrangians
\cite{Lutz01,Lutz02,Spain2,Spain5,Lutz1,Spain3,Spain4,Spain6,Garcia-Recio}
suggest that this resonance
is generated by meson-baryon interactions.

The different nature of the $\Sigma(1385)$ and of the $\Lambda(1405)$ resonances
is built in the chiral coupled-channel approach of kaon-nucleon scattering
develo\-ped in Ref. \cite{Lutz1} and which will be used in this work. The baryon
resonances belonging to the large N$_c$ ground state baryon mutiplets
(hence the $\Sigma(1385)$) are introduced explicitly as fundamental fields
of the effective Lagrangian. The other baryon resonances (in
particular the $\Lambda(1405)$) are generated dynamically by meson-baryon
coupled-channel dynamics.

The $\Sigma(1385)$ and $\Lambda(1405)$ resonances have also different spectral properties.
The $\Sigma(1385)$ mass distribution is very close to a Breit-Wigner form \cite{Barreiro}.
The spectral shape of the $\Lambda(1405)$ departs from a Breit-Wigner \cite{Hemingway,Dalitz3}.
It depends strongly on the initial and final states through which it is measured,
emphasizing the need for a full understanding of the coupling of the $\Lambda(1405)$
to its different decay channels.

We study the $\gamma p \rightarrow K^+ \pi^0 \Lambda$ and $\gamma p \rightarrow K^+ \pi \Sigma $
reactions with the idea of using future accurate data on these processes (mainly
t-distributions) to gain understanding of the $K^- p \rightarrow \pi^0 \Lambda$
and of the  $K^- p \rightarrow \pi \Sigma$ amplitudes below the $\bar K N$ threshold,
where they are dominated by the $\Sigma(1385)$ and $\Lambda(1405)$ re\-sonances.
This procedure requires that these reactions be significantly driven by the process
in which the ingoing photon dissociates
into a real K$^+$ and a virtual K$^-$, the off-shell K$^-$ scattering subsequently
off the proton target to produce the $\pi^0\, \Lambda$ or $\pi\, \Sigma$ pair.
Such dynamics would show in a sharp drop of the differential cross sections $d\sigma/dt$
with increasing $|t|$. This drop has both a double pole component behaving like 1/$(m_K^2-t)^2$
and a single pole dependence going like 1/$(m_K^2-t)$. The amplitude associated with the
K$^-$ t-channel exchange alone is not gauge-invariant. There are many other terms of
order $\alpha$ building up the gauge-invariant amplitude. We argue that the contributions
from all these other terms to the cross section will not affect the double pole term,
which is made gauge-invariant and entirely fixed by our calculation. This issue is of interest
because the double kaon pole term contribution to the $\gamma p \rightarrow K^+ \pi\, Y$
cross section is large (contrary to what is found in other meson photoproduction
processes as discussed below). Our hope is that very accurate data
will make it possible to isolate this term by
expressing the differential cross sections $d\sigma/dt$ as a superposition of
double and single K$^-$ pole terms with less singular contributions.
Such an information would be most
useful in constraining $\bar K N$ dynamics at low and subthreshold energies.
The $K^- p$ inelastic cross sections
close to threshold are indeed poorly known \cite{Lutz1}.

We compute the t-channel K$^-$-exchange contribution to the $\gamma p \rightarrow K^+ \pi^0 \Lambda$ and
$\gamma p \rightarrow K^+ \pi \Sigma $
reactions for photon laboratory energies around 2 GeV using the $K^- p \rightarrow \pi^0 \Lambda$
and $K^- p \rightarrow \pi \Sigma$ amplitudes obtained
in Ref. \cite{Lutz1} and extract the gauge-invariant double $K^-$ pole contribution for both processes.
 The steps of that calculation
are outlined in Section 2. Our numerical results for the
$\gamma \, p \rightarrow K^+ \, \pi^0 \,\Lambda(1116)$ and
$\gamma\, p \rightarrow K^+\, \pi\, \Sigma(1192) $ reactions based on the double K$^-$ pole term
are displayed
and discussed in Section 3. We conclude by a few remarks in Section 4.

\section{The $\gamma \, p \rightarrow K^+ \, \pi^0 \,\Lambda(1116)$ and
$\gamma\, p \rightarrow K^+\, \pi\, \Sigma(1192) $ reaction cross sections}

The t-channel K$^-$-exchange
contributions
to the amplitudes for the  $\gamma \, p \rightarrow K^+ \,
\pi^0 \,\Lambda(1116)$ and
$\gamma\, p \rightarrow K^+\, \pi\, \Sigma(1192) $ reactions are
displayed in Figs. 1 and 2.
\vskip 0.4true cm
\begin{figure}[h]
\noindent
\begin{center}
\mbox{\epsfig{file=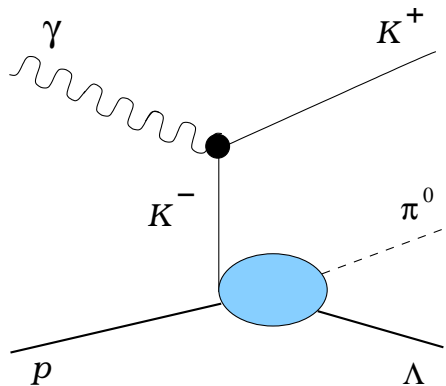, height=5 truecm}}
\end{center}
\vskip 0.4 true cm
\caption{K$^-$-exchange contribution to the
$\gamma \,p\rightarrow K^+ \pi^0 \Lambda$ amplitude.}
\label{f1}
\end{figure}
\vglue 0.4truecm
\noindent
\begin{figure}[h]
\noindent
\begin{center}
\mbox{\epsfig{file=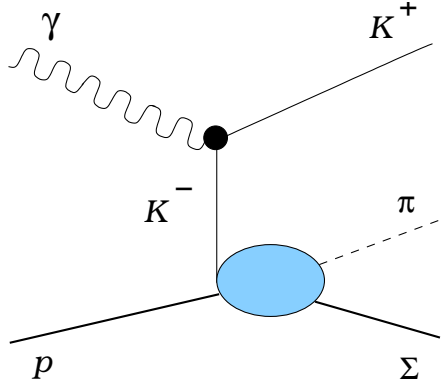, height=5 truecm}}
\end{center}
\vskip 0.4 true cm
\caption{K$^-$-exchange contribution to the
$\gamma \,p\rightarrow K^+ \pi \Sigma$ amplitude. The $\pi \Sigma$
symbol stands for $\pi^- \Sigma^+$, $\pi^0 \Sigma^0 $ or $\pi^+ \Sigma^-$.}
\label{f2}
\end{figure}
\vglue 0.3 true cm
\noindent
The importance of the contribution of the
K$^-$-exchange term to the
$\gamma \, p \rightarrow K^+ \, \pi^0 \,\Lambda(1116)$ and
$\gamma\, p \rightarrow K^+\, \pi\, \Sigma(1192) $ cross sections
can but be assessed by accurate $d\sigma/dt$ measurements for
both reactions. As discussed earlier, such data are expected in the near future
but not yet available.\par\vskip 0.2 true cm
In order to get nevertheless a rough feeling for the t-dependence of the cross section, we have
used the old data of Ref. \cite{Azemoon} characterizing the sum of the
$\gamma_v \, p \rightarrow K^+ \,\Sigma^0(1385)$ and
$\gamma_v \, p \rightarrow K^+ \,\Lambda(1405)$ processes measured
with space-like photons, in electroproduction
experiments. These data are displayed in Fig. 3.
\vglue 0.2true cm
\begin{figure}[h]
\noindent
\begin{center}
\mbox{\epsfig{file=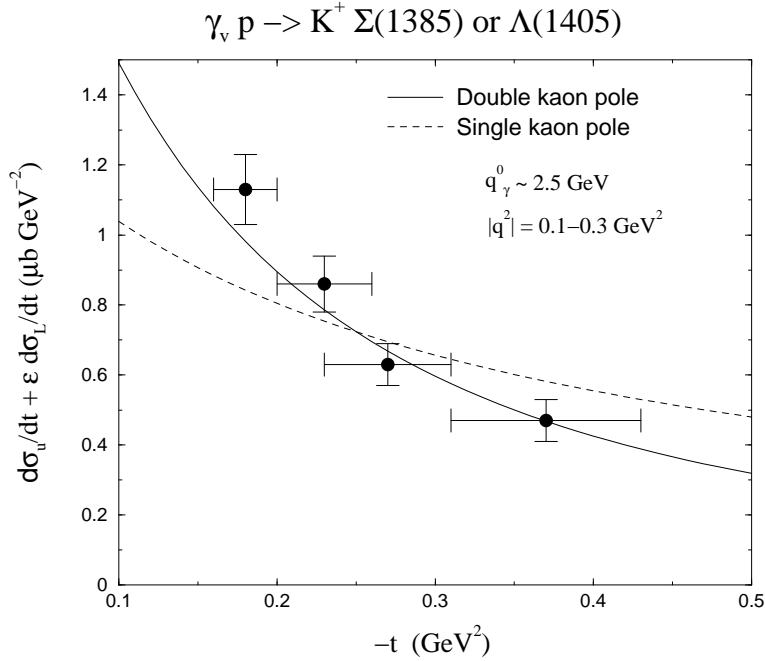, height=10 true cm, angle=-90}}
\end{center}
\vskip 0.4true cm
\caption{Differential cross section for the production of $ K^+ \Lambda$(1405) and
$ K^+ \Sigma^0$(1385) final states induced by the scattering of virtual
photons from proton targets. The data are from Ref. [5]. The quantity q$^2$ is the photon virtuality
and q$^0_\gamma$ is the (virtual) photon energy.
The two points at t=-0.27 GeV$^2$ and -0.37 GeV$^2$ are extrapolated from data
taken at higher energy according to a prescription explained in the text.
The full and dashed lines show the shape of the t-dependence of the cross section as
expected from double and single K$^-$-pole terms respectively
(with arbitrary normalizations).}
\label{f3}
\end{figure}
\par
We plot the
sum of the unpolarized transverse and longitudinal cross sections
which dominate the process ($\varepsilon$ is the transverse photon polarization) \cite{Azemoon}.
The points at
$t=-0.18$ GeV$^2$ and $t=-0.23$ GeV$^2$ are direct measurements. The points corresponding to
$t=-0.27$ GeV$^2$ and $t=-0.37$ GeV$^2$ have been measured at $q^0_\gamma\simeq3.5$ GeV and
extrapolated to $q^0_\gamma=2.5$ GeV by rescaling the diffe\-rential cross section
according to the energy dependence predicted by our model
(i.e. an increase of about a factor of 2). There is clearly a large
uncertainty associated with this procedure. We consider therefore the data displayed
in Fig. 3 only as a tenuous indication
that the reaction could proceed through a t-channel exchange in qualitative
agreement with the rapid fall-off expected from a dominant double K$^-$-pole term.\par

It should be noted however that there are no good reasons to expect significant
s-channel contributions to the $\gamma p \rightarrow K^+ \pi^0 \Lambda$ and
$\gamma p \rightarrow K^+ \pi \Sigma $ reactions at E$_\gamma\simeq2$ GeV.
The corresponding total center of mass energy in these kinematics is
$\sqrt s= 2.15$ GeV. There are no baryon resonances in that mass range
known to decay into the $K^+ \pi^0 \Lambda$ or $K^+ \pi \Sigma $ channels.
It is interesting to recall that, for the $\gamma p \rightarrow K^+ \Lambda$ and
$\gamma p \rightarrow K^+ \Sigma^0 $ reactions at E$_\gamma\simeq2$ GeV,
the dominance of the K$^-$-exchange  at low t could be inferred from photo- and
electroproduction data \cite{Guidal}. The importance of the $\Sigma^0$(1385) and
$\Lambda$(1405) resonances in the $\bar KN$ dynamics close to threshold
provides additional grounds for expecting a similar picture to hold
for the $\gamma p \rightarrow K^+ \Sigma^0 $(1385)
and $\gamma p \rightarrow K^+ \Lambda$(1405) reactions.

We calculate the cross section for the $\gamma \, p \rightarrow K^+ \, \pi \,Y$,
where Y denotes either the $\Lambda(1116)$ or the $\Sigma(1192)$. The 4-momenta
of the photon, the proton, the K$^+$, the pion and the hyperon are denoted
by $q$, $p$, $\bar q_K$, $\bar q_\pi$ and $\bar p_Y$ respectively.
The photon, proton and hyperon polarizations are indicated by the
symbols $\lambda_\gamma$, $\lambda$ and $\bar
\lambda_Y$. The total
cross section reads
\begin{eqnarray}
\sigma_{\gamma \,p\rightarrow K^+ \, \pi \,Y}&=&
\frac{1}{|{\vec v}_\gamma-{\vec v}_p|}\,\frac{1}{2\,q^0}\,\frac{m_p}{p^0}
\int \frac{d^3 \vec{\bar q}_K}{(2\pi)^3}\,\frac{1}{2\,{\bar q}\,^0_K}\,
\int \frac{d^3 \vec{\bar q}_\pi}{(2\pi)^3}\,\frac{1}{2\,{\bar q}\,^0_\pi}\,
\int \frac{d^3 \vec{\bar p}_Y}{(2\pi)^3}\,\frac{m_Y}{{\bar p}\,^0_Y}\,
\nonumber\\
&&(2\,\pi)^4\,\delta^4(q+p-{\bar q}_K-\bar q_\pi-\bar p_Y)\sum_{\lambda_\gamma,
\lambda ,\bar
\lambda_Y }\frac{1}{4}\,|M_{\gamma \,p\rightarrow K^+ \, \pi \,Y}|^2.
\label{eq1}
\end{eqnarray}
The first step of the calculation is to factorize the full amplitude
$M_{\gamma \,p\rightarrow K^+ \, \pi \,Y}$ into the
photon-kaon vertex and the $K^-\,p\rightarrow \pi \,Y$ amplitude
in accordance with the reaction mechanism depicted in Figs. 1 and 2.
We have
\begin{eqnarray}
\sum_{\lambda_\gamma,
\lambda ,\bar
\lambda_Y }\,\frac{1}{4}\,\mid M_{\gamma \,p\rightarrow K^+ \, \pi \,Y}\mid^2
&=&-e^2\,\frac{(t+m_K^2)\,}{(t-m_K^2)^2}\,\frac{1}{2}\,
\sum_{\lambda ,\bar
\lambda_Y }\,|M_{K^-p \to \pi\,Y}|^2.
\label{eq2}
\end{eqnarray}
We work in the photon-proton center of mass reference frame
where the total energy of the reaction is denoted by $\sqrt s$.
In that reference frame, the photon 3-momentum is -${\vec p}$
and the 3-momentum of the $\pi\,Y$ pair is -${\vec{\bar q}_K}$.
It is useful to define the invariant mass $\sqrt{{\bar w}^2}$ of the final $\pi\,Y$ pair by
\begin{eqnarray}
{\bar w}^2 = (p+q-{\bar q}_K)^2
 = s+ m_K^2- 2\,\sqrt{s}\,\sqrt{m_K^2+{\vec{\bar q}_K}^{\,2}} \,
\label{eq3}
\end{eqnarray}
and to express the 4-momentum transfer $t=(q-{\bar q}_K)^2$ as function of
that variable,
\begin{eqnarray}
t\;(s,{\bar w}^2,\cos \theta) &=&
 m_K^2 -\frac{1} {2s} (s-m_p^2)\, (s+m_K^2-{\bar w}^2)\,\nonumber\\
&&\mkern 110 mu \Big(1 - \sqrt{1-\frac{4\,m_K^2\,s}{(s+m_K^2-{\bar w}^2)^2}} \,\cos \theta\Big),
\label{eq4}
\end{eqnarray}
where $\theta$ is the angle between the initial photon and the produced kaon.

Using these variables, the total cross section (1) can be rewritten as
\begin{eqnarray}
\sigma_{\gamma \,p\rightarrow K^+ \, \pi \,Y} &=&
\frac{\alpha \,m_p}{16\, \pi \, s \,|\vec p\,|^2}\,
\frac{(t+m_K^2)\,}{(t-m_K^2)^2}
\,\int_{(M_Y +m_\pi)^2}^{(\sqrt{s}-m_K)^2} d {\bar w}^2
\int_{t_+(s,{\bar w}^2)}^{t_-(s,{\bar w}^2)} dt \,
\int \frac{d^3 \vec{\bar q}_\pi}{(2\pi)^3}\,\nonumber\\
&&\mkern -60 mu\frac{1}{2\,{\bar q}\,^0_\pi}\,
\int \frac{d^3 \vec{\bar p}_Y}{(2\pi)^3}\,\frac{m_Y}{{\bar p}\,^0_Y}\,
 (2\,\pi)^4\,\delta^4({\bar w}-q_\pi-\bar p_Y)\,
\frac{1}{2}\,
\sum_{\lambda ,\bar
\lambda_Y }\,|M_{K^-p \to \pi\,Y}|^2,
\label{eq5}
\end{eqnarray}
with $t_-(s,{\bar w}^2)=t\;(s,{\bar w}^2,\cos \theta=+1)$ and
$t_+(s,{\bar w}^2)=t\;(s,{\bar w}^2,\cos \theta=-1)$.

It is convenient to express Eq. (5) in terms of the total $K^-p \to \pi\,Y$
cross section, $\sigma_{K^-p \to \pi\,Y}$. This quantity is frame independent.
Its expression in the $K^-p$ center of mass reads
\begin{eqnarray}
\sigma_{K^-\,p\rightarrow \pi \,Y} &=&
\frac {1} {\sqrt{{\bar w}^2} \, |\vec q_{K^-p}|} \, \frac{m_p}{2}\,
\int \frac{d^3 \vec{\bar q}_\pi}{(2\pi)^3}\,\frac{1}{2\,{\bar q}\,^0_\pi}\,
\int \frac{d^3 \vec{\bar p}_Y}{(2\pi)^3}\,\frac{m_Y}{{\bar p}\,^0_Y}\,
 \nonumber\\
&&\mkern 90 mu(2\,\pi)^4\,\delta^4({\bar w}-\bar q_\pi-\bar p_Y)\,
\frac{1}{2}\,
\sum_{\lambda ,\bar
\lambda_Y }\,|M_{K^-p \to \pi\,Y}|^2,
\label{eq6}
\end{eqnarray}
in which $q_{K^-p}$ is the $K^-$ momentum in the $K^-p$ center of mass.
Its value as function of the total center of mass energy $\sqrt {{\bar w}^2}$
of the $K^-p$ system
is given by
\begin{eqnarray}
|\vec q_{K^-p}|^2=
\frac{1}{4\,{\bar w}^2}\, {\{{\bar w}^4-2\,{\bar w}^2\,(m_p^2+m_K^2)+(m_p^2-m_K^2)^2\}}.
\label{eq7}
\end{eqnarray}
\par
\newpage
The doubly differential cross section $d\sigma/dt d{\bar w}^2$ can be written as
\begin{eqnarray}
\frac{d\sigma_{\gamma\,p\rightarrow K^+\,\pi \,Y}} {dt \, d{\bar w}^2} &=&
\frac{\alpha}{4\, \pi}\, \frac{({\bar w}^4-2\,{\bar w}^2\,(m_p^2+m_K^2)+(m_p^2-m_K^2)^2)^{1/2}}{(s-m_p^2)^2}\,
\nonumber\\
&&\mkern 150 mu\frac{(t+m_K^2)\,}{(t-m_K^2)^2}\,\sigma_{K^-\,p\rightarrow \pi \,Y}({\bar w}^2).
\label{eq8}
\end{eqnarray}

We remark first that the amplitude $M_{\gamma \,p\rightarrow K^+ \, \pi \,Y}$
obtained by calculating the graph of Fig. 1 (or Fig. 2) is not gauge-invariant.
To obtain the full gauge-invariant amplitude, all the other diagrams of order $\alpha$ leading
to the same final state should be added.
In the energy range under consideration
(E$_\gamma\simeq2$ GeV), there are many possible diagrams involving poorly known
couplings and hence large uncertainties.

Instead of attempting to calculate these graphs, we resort to the pole scheme
method \cite{Veltman}. This technique was used to derive gauge-invariant
results in the vicinity of a pole for electroweak processes
involving radiative corrections \cite{Stuart}. The idea of the method
is to decompose the amplitude according to its pole structure and to expand
it around the pole. To any order in perturbation theory, the residues
of the poles are gauge-invariant. The expansion provides therefore
subsets of gauge-invariant expressions associated with a given pole structure.

We apply this method to derive the gauge-invariant cross section
correspon\-ding to the double K$^-$-pole term to first order in $\alpha$.
We expect this term to play a significant role in the $\gamma \,p\rightarrow K^+ \, \pi \,Y$
process. The key point is that the graph of Fig. 1 (or Fig. 2) is the only process
which can contribute to the
double K$^-$-pole term. We will therefore decompose the corresponding cross section according to
its pole structure, keep only the double K$^-$-pole term and extract the
gauge-invariant cross section associated with that pole structure by
calculating the residue at the pole.

According to this procedure, the gauge-invariant cross section
corresponding to the double K$^-$-pole term reads
\begin{eqnarray}
\frac{d\sigma_{\gamma\,p\rightarrow K^+\,\pi \,Y}} {dt\, d{\bar w}^2} &=&
\frac{\alpha}{2\, \pi}\, \frac{({\bar w}^4-2\,{\bar w}^2\,(m_p^2+m_K^2)+(m_p^2-m_K^2)^2)^{1/2}}{(s-m_p^2)^2}\,
\nonumber\\
&&\mkern 150 mu\frac{m_K^2\,}{(t-m_K^2)^2}\,\sigma_{K^-\,p\rightarrow \pi \,Y}({\bar w}^2).
\label{eq9}
\end{eqnarray}
\par
We stress that the double pole term is the only one which can be determined
this way, because it does not get contributions from any other graph but the
t-channel kaon-exchange diagram.
In contrast, the single pole term is built up as a sum of many processes, in particular
the interference of the t-channel kaon-exchange diagram with s- and
u-channel terms.

In order to be able to extract the double pole term from accurate
t-distributions, it has to be reasonably large. We speculate so in
view of the numerical results discussed in the next section. As shown
earlier, it is also compatible with the few data points available.

We emphasize that this would be a specific behaviour of the
${\gamma\,p\rightarrow K^+\,\pi \,Y}$ reactions. The situation is indeed
different in other meson photoproduction processes, even when they are
dominated by t-channel exchanges.
It is inte\-resting to consider for example the pole structure for the photoproduction
of $\omega$-mesons in the energy range $1.4<E_\gamma<1.6$ GeV. In these
kinematics, the $\gamma p \rightarrow \omega p$ reaction is expected to be well
described by a simple pion-exchange diagram for small momentum transfers
 and with standard form factors \cite{Friman}.
We use the specific model of Ref. \cite{Friman}. Neglecting the t-dependence of
the form factors and concentrating on the pole structure of the differential
cross section in the limit of local couplings, we have\par
\begin{eqnarray}
\frac{d\sigma_{\gamma\,p\rightarrow \omega \,p}} {dt} &\propto&
\frac{-t}{4\, m_p^2}\, \frac{(m_{\omega}^2-t)^2}{(m_{\pi}^2-t)^2}\nonumber\\
&&=\frac{-m_\pi^2}{4\, m_p^2}\,\frac{(m_\omega^2-m_\pi^2)^2}
{(m_\pi^2-t)^2}+\frac{(m_\omega^2-m_\pi^2)}{4\, m_p^2}\,\frac{(m_\omega^2-3m_\pi^2)^2}
{(m_\pi^2-t)}\nonumber\\&&\mkern 184mu+\frac{1}{4\, m_p^2}\,{(2m_\omega^2-2m_\pi^2-t)}.
\label{eq10}
\end{eqnarray}
We see that the differential cross section behaviour is
dominated by the single pion pole term. The double pion pole is small
and even negative. It acts as a minor correction to the cross section,
as a consequence of the smallness of the pion mass. The pole scheme
technique used above is not predictive in this case and the
dominant single pion pole term has to be determined phenomenologically.
As a consequence of this observation, the effective
$\pi NN$ and $\pi \gamma \omega$ couplings are not constrained
directly by the data and depend on further assumptions such as the
form factors assigned to the vertices.
We remark that the particular form of the couplings chosen
in Ref. \cite{Friman}
makes the full pion-exchange diagram gauge-invariant, irrespectively of its pole structure.
\par
To characterize the momentum transfer dependence of the $K^-\,p\rightarrow \pi \,Y$
cross section, one can define the variable $\bar t\,=\, (p-{\bar p}_Y)^2$ and generalize Eq.(9)
to the threefold
differential cross section
\newpage
\begin{eqnarray}
\frac{d\sigma_{\gamma\,p\rightarrow K^+\,\pi \,Y}} {dt \, d{\bar w}^2 d{\bar t}} &=&
\frac{\alpha}{2\, \pi}\, \frac{({\bar w}^4-2\,{\bar w}^2\,(m_p^2+m_K^2)+(m_p^2-m_K^2)^2)^{1/2}}{(s-m_p^2)^2}\,
\nonumber\\
&&\mkern 120 mu\frac{m_K^2\,}{(t-m_K^2)^2}\,\frac{d\sigma_{K^-\,p\rightarrow \pi \,Y}}
{d{\bar t}}({\bar w}^2,\,{\bar t}).
\label{eq11}
\end{eqnarray}
\par
We focus on the ${K^-\,p\rightarrow \pi \,Y}$ reaction by integrating over t, using the
kinematic boundaries $t_-$ and $t_+$ defined after Eq. (5). We have
\begin{eqnarray}
\int_{t_+}^{t_-} dt \,
\frac{d\sigma_{\gamma\,p\rightarrow K^+\,\pi \,Y}} {dt \, d{\bar w}^2 d{\bar t}} &=&
\frac{\alpha}{2\, \pi}\, \frac{({\bar w}^4-2\,{\bar w}^2\,(m_p^2+m_K^2)+(m_p^2-m_K^2)^2)^{1/2}}{(s-m_p^2)^2}\,
\nonumber\\
&&\mkern 35 mu\frac{m_K^2\,(t_--t_+)}{(t_+-m_K^2)\,(t_--m_K^2)}\,\frac{d\sigma_{K^-\,p\rightarrow \pi \,Y}}
{d{\bar t}}({\bar w}^2,\,{\bar t}),
\label{eq12}
\end{eqnarray}
\par
or\par
\begin{eqnarray}
\frac{d\sigma_{\gamma\,p\rightarrow K^+\,\pi \,Y}} {d{\bar w}^2 d{\bar t}} &=&
\frac{\alpha}{2\,\pi}\,\frac{{|\vec{\bar q}_K}|\,s^{1/2}}{(s-m_p^2)^3}\,
4\,|\vec q_{K^-p}|\sqrt{{\bar w}^2}\,
\frac{d\sigma_{K^-\,p\rightarrow \pi \,Y}}
{d{\bar t}}({\bar w}^2,\,{\bar t}).
\label{eq13}
\end{eqnarray}
\par
The nontrivial dynamical quantity of interest in Eq. (13) is clearly
$4\,|\vec q_{K^-p}|\sqrt{{\bar w}^2}$\\
\noindent$\frac{d\sigma_{K^-\,p\rightarrow \pi \,Y}}{d{\bar t}}({\bar w}^2,\,{\bar t})$.

To calculate this cross section we use the $\bar K p\rightarrow \pi Y$ amplitudes derived
in Ref. \cite{Lutz1} from the chiral SU(3) Lagrangian by solving coupled-channel
Bethe-Salpeter equations. We will not repeat here the technical developments involved
in this scheme. They are explained and discussed extensively in Ref. \cite{Lutz1}.

This effective field theory achieves an excellent description of the available data
on $K^-\,p$ elastic (direct and charge-exchange) and inelastic ($\pi^0\,\Lambda$,
$\pi^+\,\Sigma^-$, $\pi^0\,\Sigma^0$, $\pi^-\,\Sigma^+$) processes up to
laboratory $K^-$ momenta of the order of 500 MeV. The interest of the
present work is to offer the possibility of testing the amplitudes
below the $\bar K N$ threshold, in the region
where they are dominated by the $\Lambda$(1405) and the $\Sigma$(1385).
The specific spectral shape of these resonances is a particularly
meaningful prediction
of the description of Ref. \cite{Lutz1}. As mentioned earlier, the shape of the $\Sigma$(1385)
resonance is expected to be close to a Breit-Wigner form. The $\Lambda$(1405)
resonance mass distribution has an asymmetric shape
and depends on the initial and final states through which it is observed.
The $\pi^0\,\Lambda$ production would also put strong constraints on the
$\Sigma$(1385)$\, \bar K N$ coupling which is poorly known.

\newpage
\section{Numerical results}
We present our numerical results in three different perspectives. We restrict our
calculations to cross sections integrated over $\bar t$.

We show first the quantity $4\,|\vec q_{K^-p}|\sqrt{{\bar w}^2}$
$\sigma_{K^-\,p\rightarrow \pi \,Y}({\bar w}^2)$
as function of the total center of mass energy in the $K^- p$ system, renamed for clarity
$\sqrt{s_{K^-p}}$ ($\equiv{\sqrt{{\bar w}^2}}$). The interest of displaying our results
this way is to exhibit the behaviour of the ${K^-\,p\rightarrow \pi \,Y}$ cross section
across threshold. We recall that the $\bar K N$ threshold is at $\sqrt{s_{K^-p}}\,\approx\,$1.435 GeV.
Our predicted cross sections  are displayed in Fig. ~4 for the ${K^-\,p\rightarrow \pi^- \,\Sigma^+}$ and
${K^-\,p\rightarrow \pi^+ \,\Sigma^-}$ reactions and in Fig. 5 for the
${K^-\,p\rightarrow \pi^0 \,\Sigma^0}$ and
${K^-\,p\rightarrow \pi^0 \,\Lambda}$ reactions. They are compared to the data
available above threshold \cite{Mast1,Sakitt,Ciborowski,Bangerter,Armenteros,Humphrey}.
\begin{figure}[h]
\noindent
\begin{center}
\mbox{\epsfig{file=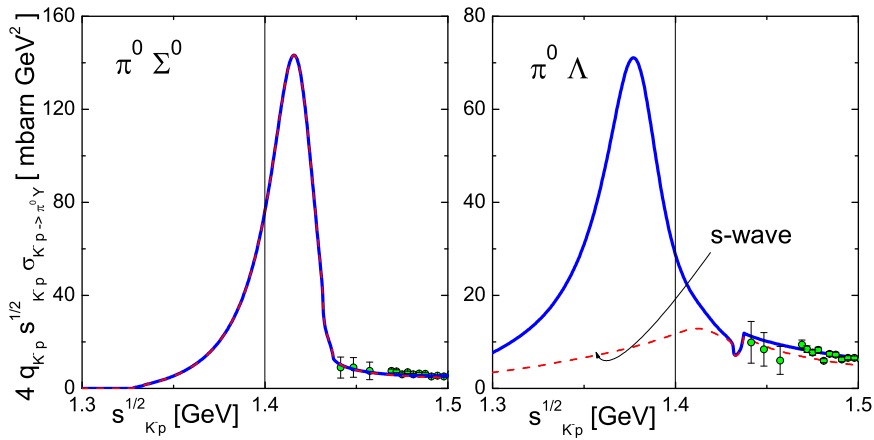, height=7 true cm}}
\end{center}
\vskip 0.4 true cm
\caption{
$K^- \, p \rightarrow \pi^- \,\Sigma^+$ and $K^- \, p \rightarrow \pi^+ \,\Sigma^-$
cross sections below and above threshold. The length of the $K^-$ 3-momentum is defined
by Eq. (7) and $\sqrt{s_{K^-p}}$ ($\equiv{\sqrt{{\bar w}^2}}$) is the total center of mass energy
of the  $K^- \, p$ system. The dashed line represents the contribution from $K^- \, p$
relative s-wave only. The grey histogram is explained in the text.
The data above threshold are from Refs. \cite{Mast1,Sakitt,Ciborowski,Bangerter,Armenteros,Humphrey}.}
\label{f4}
\end{figure}
\par
We recall that the $\pi \,\Sigma$ channel is dominated by the $\Lambda$(1405)
and the $\pi \,\Lambda$ channel by the $\Sigma$(1385).
The properties of the spectral functions of the $\Sigma$(1385) and $\Lambda$(1405)
resonances are very apparent in Figs. 4 and 5. The shape of the resonant behaviour
of the ${K^-\,p\rightarrow \pi^0 \,\Lambda}$ cross section below threshold is quite symmetric and close
to a Breit-Wigner form. The s-wave contribution is small as expected for a
process dominated by a p-wave
resonance. In contrast, the spectral form of the ${K^-\,p\rightarrow \pi \,\Sigma}$
cross sections
for the three possible $\pi \,\Sigma$ channels is asymmetric and largely given by
s-wave dynamics, reflecting the $\Lambda$(1405) dominance.
The grey histograms show (in arbitrary units) the empi\-rical shape
of the $\Lambda(1405)$ resonance extracted from the $p\,(\gamma, K^+
\pi)\, \Sigma$ reaction at 1.5-2.4 GeV photon energy \cite{Ahn}. It
should be emphasized that strictly speaking the comparison implied
by Fig. 4 is not yet justified. Only after extracting the double
kaon pole contribution from the cross section by a detailed study of
t-distributions can this comparison become legitimate. The result of such an
analysis is expected to resolve the discrepancy of the histograms
and the $K^-p$ scattering data, at least above threshold.
Nevertheless, the different line shapes of \cite{Ahn} seems to
confirm the prediction of chiral coupled-channel dynamics that the
spectral shape of the $\Lambda(1405)$ resonance depends crucially
on the initial and final states it is probed with
\cite{Lutz1,Nacher}. We note that the available
$K^-p$ scattering data close to threshold have large error bars,
emphasizing the interest of being able to determine from
experiment the subthreshold $K^-p$ scattering amplitudes by
extracting the double kaon pole contributions to the $\gamma p \to
K^+ \pi Y$ reactions calculated in this work.
\par
\begin{figure}[h]
\noindent
\begin{center}
\mbox{\epsfig{file=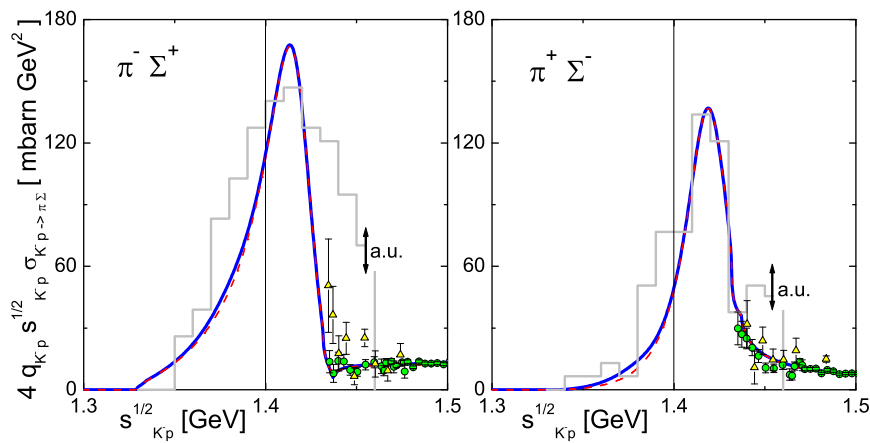, height=7 true cm}}
\end{center}
\vskip 0.4 true cm
\caption{Same as Fig. 4 for the
$K^- \, p \rightarrow \pi^0 \,\Sigma^0$ and $K^- \, p \rightarrow \pi^0 \,\Lambda$ channels.
}
\label{f5}
\end{figure}
\par
We display in Figs. 6 and 7 the
double kaon pole term contributions to the
differential cross sections for the $\gamma p \rightarrow K^+ \pi^0 \Lambda$ and
$\gamma p \rightarrow K^+ \pi \Sigma $ reactions
as functions of the $\pi\, Y$ total center of mass energy $\sqrt{s_{K^-p}}$ at $E_\gamma=1.7$ GeV
and $E_\gamma=2.1$ GeV.
\begin{figure}[h]
\noindent
\begin{center}
\mbox{\epsfig{file=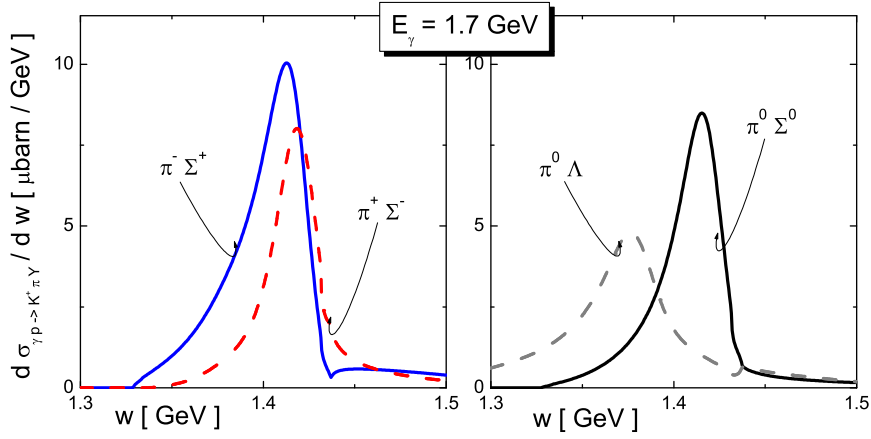, height=7 true cm}}
\end{center}
\vskip 0.4 true cm
\caption{Double kaon pole term contribution to the
differential cross sections for the $\gamma p \rightarrow K^+ \pi^0 \Lambda$ and
$\gamma p \rightarrow K^+ \pi \Sigma $ reactions
as function of the $\pi\, Y$ total center of mass energy at $E_\gamma=1.7$ GeV
}
\label{f6}
\end{figure}
\par
\begin{figure}[h]
\noindent
\begin{center}
\mbox{\epsfig{file=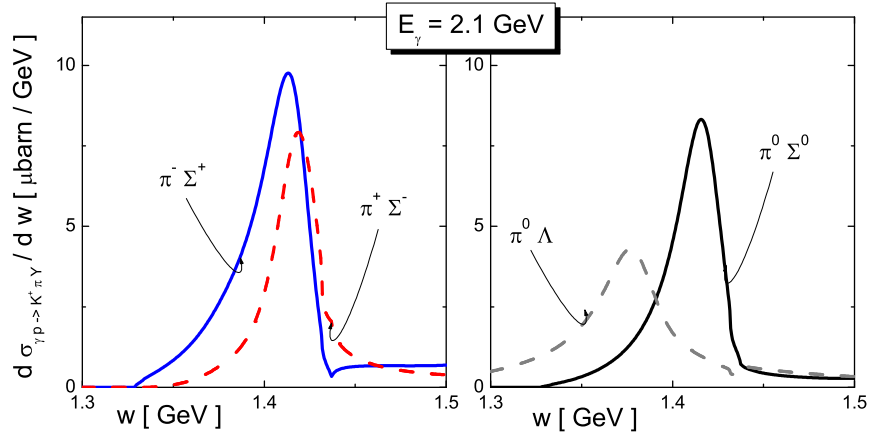, height=7 true cm}}
\end{center}
\vskip 0.4 true cm
\caption{Same as Fig. 6 at $E_\gamma=2.1$ GeV
}
\label{f7}
\end{figure}
\par
They reflect clearly the dynamical features discussed in commenting on
Figs. 4 and 5. It is also interesting to note the absolute values of the
double kaon pole cross sections.
They are large on the scale of what is expected from other theoretical approaches.
If we compare our results to the predictions of the model
of Ref. \cite{Nacher} at $E_\gamma=1.7$ GeV, we notice that our calculated
cross sections are roughly twice larger for the $\pi\,\Sigma$
channels.
It is not easy to trace the origin of this effect. Our gauge-invariant
double kaon pole term contains contributions which cannot be mapped easily
onto the Feynman diagrams computed in Ref. \cite{Nacher}.
The cross section we obtain for the $\pi^0\,\Lambda$ channel is about
an order of magnitude larger than the result displayed
in Ref. \cite{Nacher}. A substantial part of this effect should be
ascribed to the neglect of the $\Sigma$(1385) resonance in that work.
\par

Finally, we show in Fig. 8 the t-distribution of the double kaon pole term contribution
to the $\gamma p \rightarrow K^+ \pi^0 \Sigma^0$ reaction
at $E_\gamma=2.1$ GeV,
computed at different values of the $\pi^0\,\Sigma^0 $ total center of mass energy.
As mentioned in the introduction, this process is
a unique signature of the $\Lambda(1405)$ resonance.
 It is a very nice test of the underlying dynamics
of the $\Lambda(1405)$ photoproduction and a clean process to extract
the double kaon pole from accurate data.
\begin{figure}[h]
\noindent
\begin{center}
\mbox{\epsfig{file=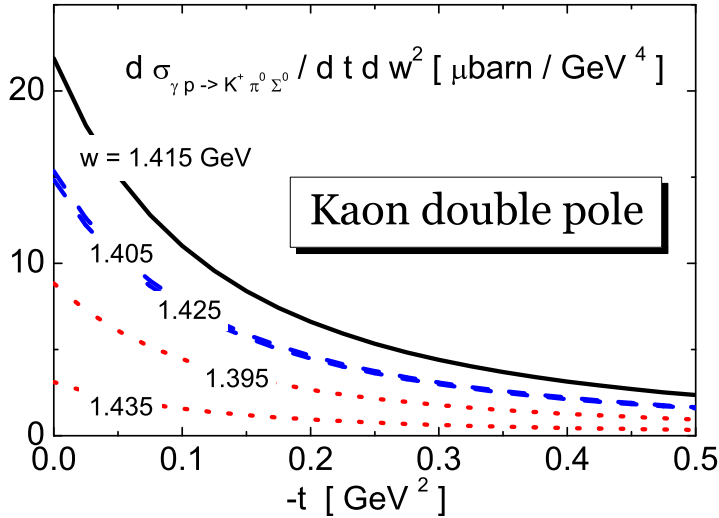, height=8 true cm}}
\end{center}
\vskip 0.4 true cm
\caption{Double kaon pole term contribution to the t-distribution
for the $\gamma p \rightarrow K^+ \pi^0 \Sigma^0$ reaction
at $E_\gamma=2.1$ GeV,
 computed at different values of the $\pi^0\,\Sigma^0 $ total center of mass energy.
}
\label{f8}
\end{figure}
\par

\section{Conclusion}
We have studied the $\gamma p \rightarrow K^+ \pi^0 \Lambda$ and $\gamma p \rightarrow K^+ \pi \Sigma $
reactions in the kinematic region where the $\pi^0 \Lambda$(1116) and $\pi\Sigma$(1192) pairs
originate dominantly from the decay of the $\Sigma$(1385) and $\Lambda$(1405) resonances.
We focus on laboratory photon energies around 2 GeV, significantly above the threshold
for producing the $K^+\,\Sigma$(1385)
(E$^{thresh}_\gamma$=1.41 GeV)
and the $K^+\,\Lambda$(1405)
(E$^{thresh}_\gamma$=1.45 GeV)
final states. We have calculated the t-channel $K^-$-exchange contribution to these reactions
using the ${K^-\,p\rightarrow \pi \,Y}$ amplitudes of Ref. \cite{Lutz1}, which
have been shown to describe the data available at low kaon momentum. Based on the pole
structure of this contribution, we determined the gauge-invariant double kaon pole contribution to the
$\gamma p \rightarrow K^+ \pi Y$ cross sections by calculating the residue at the pole.
The relevance of our work stems from the advent of detector systems able
to measure exclusively multiparticle final states with great accuracy.
Three complementary experiments in the photon energy range considered in this paper
are planned with LEPS at SPring-8 \cite{Ahn}, SAPHIR at ELSA \cite{SAPHIR}
and  the Crystal Barrel at ELSA \cite{Schmieden},
dealing for the first two with the charged [$\pi^- \Sigma^+$ and $\pi^+ \Sigma^-$] channels
and for the latter with the neutral [$\pi^0 \Sigma^0$ and $\pi^0 \Lambda$] final states.
These accurate measurements
should make it possible to extract the contribution of the double kaon pole
and hence to study kaon-nucleon dynamics
below threshold.

\section{Acknowledgement}
We acknowledge very stimulating discussions with Hartmut Schmieden. One of
us (M.S) is indebted to the hospitality of the GSI Theory Group, where part
of this work was done.


\begin{thebibliography}{99}
\bibitem{Hagiwara}K. Hagiwara et al. (Particle Data Group), Phys. Rev. D66 (2002) 010001.
\bibitem{Schmieden}H. Schmieden, Private communication.
\bibitem{Ahn}K. Ahn, Nucl. Phys. A721 (2003) 715c.
\bibitem{SAPHIR}SAPHIR Collaboration, Private communication.
\bibitem{Azemoon}T. Azemoon et al., Nucl. Phys. B 95 (1975) 77.
\bibitem{Isgur}N. Isgur, G. Karl, Phys. Rev. D18 (1978) 4187.
\bibitem{Glozman}L.Ya. Glozman, D.O. Riska, Phys. Rep. 268 (1996) 263.
\bibitem{Jaffe}R.L. Jaffe, Topical Conf. on Baryon Resonances, Oxford (England), July 5-9, 1976,
Proceedings p. 455.
\bibitem {deSwart}J.J. de Swart, P.J. Mulders, L.J. Somers,
4th Int. Conf. on Baryon Resonances, Toronto (Canada), July 14-16, 1980,
Proceedings, p. 405.
\bibitem{Kaxiras}E. Kaxiras, E.J. Moniz, M. Soyeur, Phys.Rev. D 32 (1985) 695.
\bibitem{Dalitz1}R.H. Dalitz, S.F. Tuan, Phys. Rev. Lett. 2 (1959) 425.
\bibitem{Ball}J.S. Ball, W.R. Frazer, Phys. Rev. Lett. 7 (1961) 204.
\bibitem{Arnold}R.C. Arnold, J.J. Sakurai, Phys. Rev. 128 (1962) 2808.
\bibitem{Wyld1}H.W. Wyld, Phys. Rev. 155 (1967) 1649.
\bibitem{Dalitz2}R.H. Dalitz, T.C. Wong, G. Rajasekaran, Phys. Rev. 153 (1967) 1617.
\bibitem{Logan}R.K. Logan and H.W. Wyld, Phys. Rev. 158 (1967) 1467.
\bibitem{Rajasekaran}G. Rajasekaran, Phys. Rev. D 5 (1972) 610.
\bibitem{Siegel}P.B. Siegel, W. Weise, Phys. Rev. C 38 (1988) 2221.
\bibitem{Callan}C.G. Callan, K. Hornbostel, I. Klebanov, Phys. Lett. B 202 (1988) 269.
\bibitem{Blom}U. Blom, K. Dannbom, D.O. Riska, Nucl. Phys. A 493 (1989) 384.
\bibitem{Veit1}E.A. Veit et al., Phys. Lett. B 137 (1984) 415.
\bibitem{Veit2}E.A. Veit et al., Phys. Rev. D 31 (1985) 1033.
\bibitem{Lutz01}M.F.M. Lutz, E.E. Kolomeitsev, Proc. Int. Workshop XXVIII on
Gross Properties of Nuclei and Nuclear Excitations, Hirschegg (Austria), January 16-22, 2000.
\bibitem{Lutz02}M.F.M. Lutz, E.E. Kolomeitsev, Found. Phys. 31 (2001) 1671.
\bibitem{Spain2}E. Oset, A. Ramos, C. Bennhold, Phys. Lett. B 527 (2002) 99.
\bibitem{Spain5}D. Jido et al, Phys. Rev. C 66 (2002) 055203.
\bibitem{Lutz1}M.F.M. Lutz, E.E. Kolomeitsev, Nucl. Phys. A 700 (2002) 193.
\bibitem{Spain3}C. Garcia-Recio et al., Phys. Rev. D 67 (2003) 076009.
\bibitem{Spain4}D. Jido et al, Nucl. Phys. A 525 (2003) 181.
\bibitem{Spain6}T. Hyodo et al. Phys. Rev. C 68 (2003) 065203.
\bibitem{Garcia-Recio}C. Garcia-Recio, M.F.M. Lutz, J. Nieves, Phys. Lett. B 582 (2004) 49.
\bibitem{Barreiro}F. Barreiro et al., Nucl. Phys. B 126 (1977) 319.
\bibitem{Hemingway}R.J. Hemingway, Nucl. Phys. B 253 (1985) 742.
\bibitem{Dalitz3}R.H. Dalitz, A. Deloff, J. Phys. G 17 (1991) 289.
\bibitem{Guidal}M. Guidal, J.-M. Laget, M. Vanderhaeghen, Phys. Rev. C 61 (2000) 025204.
\bibitem{Veltman}M. Veltman, Physica 29 (1963) 186.
\bibitem{Stuart}R.G. Stuart, Phys. Lett. B 262 (1991) 113.
\bibitem{Friman}B. Friman, M. Soyeur, Nucl. Phys. A 600 (1996) 477.
\bibitem{Mast1}T.S. Mast et al., Phys. Rev. D 11 (1975) 3078.
\bibitem{Sakitt}M. Sakitt et al., Phys. Rev. 139 (1965) B719.
\bibitem{Ciborowski}J. Ciborowski et al., J. Phys. G 8 (1982) 13.
\bibitem{Bangerter}R.O. Bangerter et al., Phys. Rev. D 23 (1981) 1484.
\bibitem{Armenteros} R. Armenteros et al., Nucl. Phys. B 21 (1970) 15.
\bibitem{Humphrey}W.E. Humphrey, R.R. Ross, Phys. Rev. 127 (1962) 1305.
\bibitem{Nacher}J.C. Nacher et al., Phys. Lett. B 455 (1999) 55.

\end{thebibliography}
\end{document}